\documentclass[aps,prb,preprint,superscriptaddress,nobibnotes]{revtex4}
\usepackage{graphicx}
\begin{document}

\title{Quantum strain sensor with a topological insulator HgTe quantum
  dot}

\author{Marek Korkusinski}
\email[Correspondence to ]{Marek.Korkusinski@nrc-cnrc.gc.ca}
\affiliation{Quantum Theory Group, Security and Disruptive
Technologies Portfolio, Emerging Technologies Division,
National Research Council, Ottawa, Canada K1A 0R6}

\author{Pawel Hawrylak}
\affiliation{Quantum Theory Group, Security and Disruptive
Technologies Portfolio, Emerging Technologies Division,
National Research Council, Ottawa, Canada K1A 0R6}

\date{\today}

\begin{abstract}
We present a theory of electronic properties of HgTe quantum dot 
and propose a strain sensor based on a strain-driven
transition from a HgTe quantum dot with inverted bandstructure and
robust topologically protected quantum edge states to a normal 
state without edge states in the energy gap. 
The presence or absence of edge states leads to large on/off ratio of
conductivity across the quantum dot, tunable by adjusting the number of 
conduction channels in the source-drain voltage window. 
The electronic properties of a HgTe quantum dot as a function of size and 
applied strain are described using eight-band $\vec{k}\cdot\vec{p}$
Luttinger and Bir-Pikus Hamiltonians, with surface states  identified
with chirality of Luttinger spinors and obtained through extensive
numerical diagonalization of the Hamiltonian. 
\end{abstract}

\maketitle

There is currently significant interest in using quantum effects to
develop capabilities in sensing at the nanoscale,  
from single-electron charge detection using quantum point
contacts,\cite{elzerman_hanson_nature2004,granger_taubert_natphys2012}
few nuclear spins using the quantum states of NV center in
diamond\cite{balasubramanian_chan_nature2008,mamin_kim_science2013} 
to piezotronic sensors 
of strain\cite{wang_zhou_nanolett2006,wang_advmat2012} and development
of smart skin.\cite{wu_wen_science2013} 
The piezotronic sensors rely on inducing a charge on the
surface of piezoelectric semiconductor nanowire by strain, which in
turn continuously changes the conductivity of the
nanowire.\cite{wang_zhou_nanolett2006} 
Here we describe a novel strain sensor based on a strain-driven
transition from a HgTe quantum dot with inverted bandstructure and
robust topologically protected quantum edge states to a normal 
state without edge states in the energy gap. 
The presence/absence of edge states is expected to lead to large
on/off ratio of conductivity across the quantum dot in analogy to
recently demonstrated large piezoresistive current on/off ratios 
driven by a pressure-induced metal/insulator transition in rare earth
chalcogenide thin films.\cite{copel_kuroda_nl2013}
The electronic properties of a HgTe quantum dot as a function of 
applied strain are described using eight-band $\vec{k}\cdot\vec{p}$
theory and surface states are identified with chirality of Luttinger
spinors.

\begin{figure}
\includegraphics[width=0.8\textwidth]{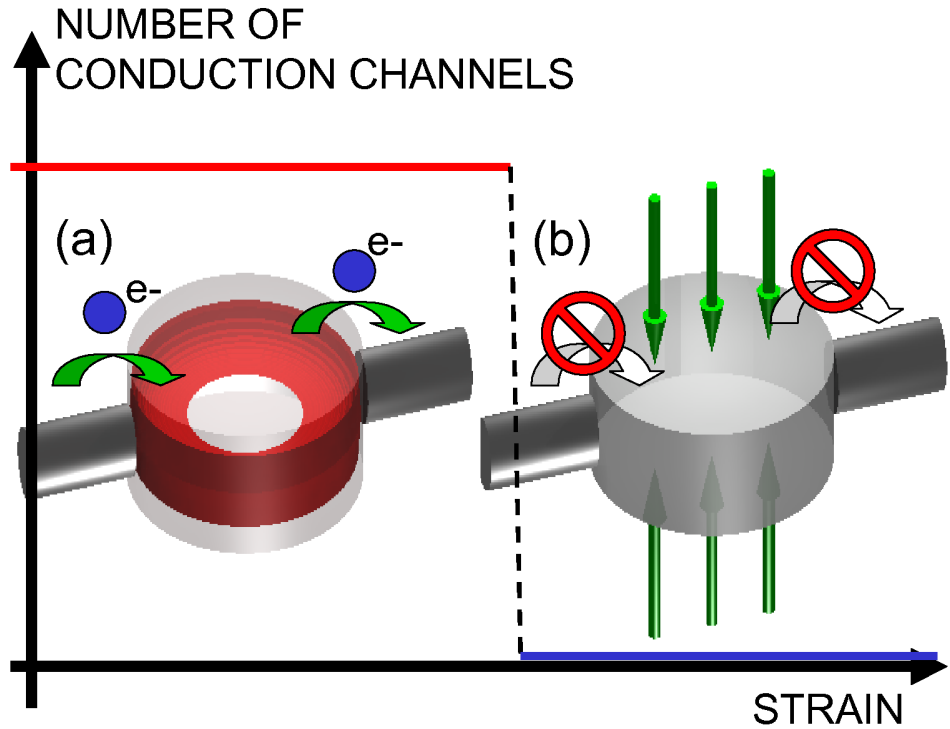}
\caption{Schematic explanation of the principle of operation of the
  strain sensor.
(a) The unstrained HgTe disk in the inverted regime is placed between
two metallic electrodes.
The Fermi energy of the electrodes matches the energy of one of the
edge states.
The current flows since the electrons tunnel easily into and out of
the edge state in the disk.
(b) Upon applied vertical stress, the gap opens in the single-particle
spectrum of the disk.
There are no states matched with the Fermi energy of the leads, the
current does not flow.
}
\label{fig1n}
\end{figure}

The principle of operation of the quantum strain sensor based on a
HgTe topological insulator quantum dot is summarized in
Fig.~\ref{fig1n}. 
Panel (a) shows a thick, finite-size HgTe quantum well in the form of
a disk embedded in vacuum or a higher bandgap material such as CdTe. 
The quantum dot is connected to two metallic electrodes. 
The quantum well of HgTe is an example of a topological insulator
(TI), a material with an energy gap in the bulk, accompanied by
helical, topologically protected  states at its
edge.\cite{hasan_kane_rmp2010,moore_nature2010,qi_zhang_rmp2011,konig_wiedmann_science2007,buettner_liu_natphys2011,knez_du_prl2012}. 
This is in contrast with normal semiconductor (NS) quantum dot, in 
which the states from the electron and hole subbands are separated by
an empty energy gap.\cite{michler-book}
The thickness of the HgTe quantum well, through quantum confinenement,
controls the relative position of the $s$ and $p$ bands.
As a result, there exists a transition from the inverted to normal
bandstructure when the thickness of the quantum well is reduced below
the critical thickness $H_C$. 
At this special thickness, the quasiparticle in-plane dispersion
correponds to a single Dirac cone.\cite{buettner_liu_natphys2011}
The thickness of the disk in Fig.~\ref{fig1n} is larger than $H_C$,
which results in the inverted bandstructure and the existence of
states, whose energy falls within the gap and probability density
peaks at the disk edge.\cite{chang_lou_prl2011}

In Fig.~\ref{fig1n}(a), the red ring  represents the computed probability
density of one of the edge states.
If the Fermi energy of the leads is aligned with the energy of the
edge state, electrons are expected to tunnel efficiently from the
right to the left lead via the edge state and one detects a high
current flow. 
Assuming ballistic transport, the current $I$ is proportional to the
number  $M$ of edge states and the applied voltage $V_{SD}$, 
$I=G_0 M T V_{SD}$,  where $G_0$  is a quantum of conductance,
$G_0=e^2/h$, and $T$ is the transmission
coefficient.\cite{buttiker_landauer,delgado_hawrylak_jphys2008,shim_delgado_prb2009} 
This is schematically visualized as the red line - high current or
large number of edge states - in Fig.~\ref{fig1n}, where we plot the
tunneling current as a function of the strain.
As we apply strain to our disk, the edges of both $s$ and $p$ bands
shift in energy, but with different deformation potentials.
If the strain is large enough, the HgTe quantum well bandstructure
is no longer inverted and the edge states disappear from the energy
gap and are out of the resonance with the Fermi levels of the leads.
Since there is no state available for the electrons to tunnel through,
the current  is expected to vanish, as depicted in Fig.~\ref{fig1n} (b).
As shown in the main panel, the vanishing of current (blue line) is
expected to occur when the  strain exceeds the critical value. 
The robustness and sharpness of edge states is expected to lead to a
very high on-off ratio of current as a function of applied strain.  

\begin{figure}
\includegraphics[width=0.8\textwidth]{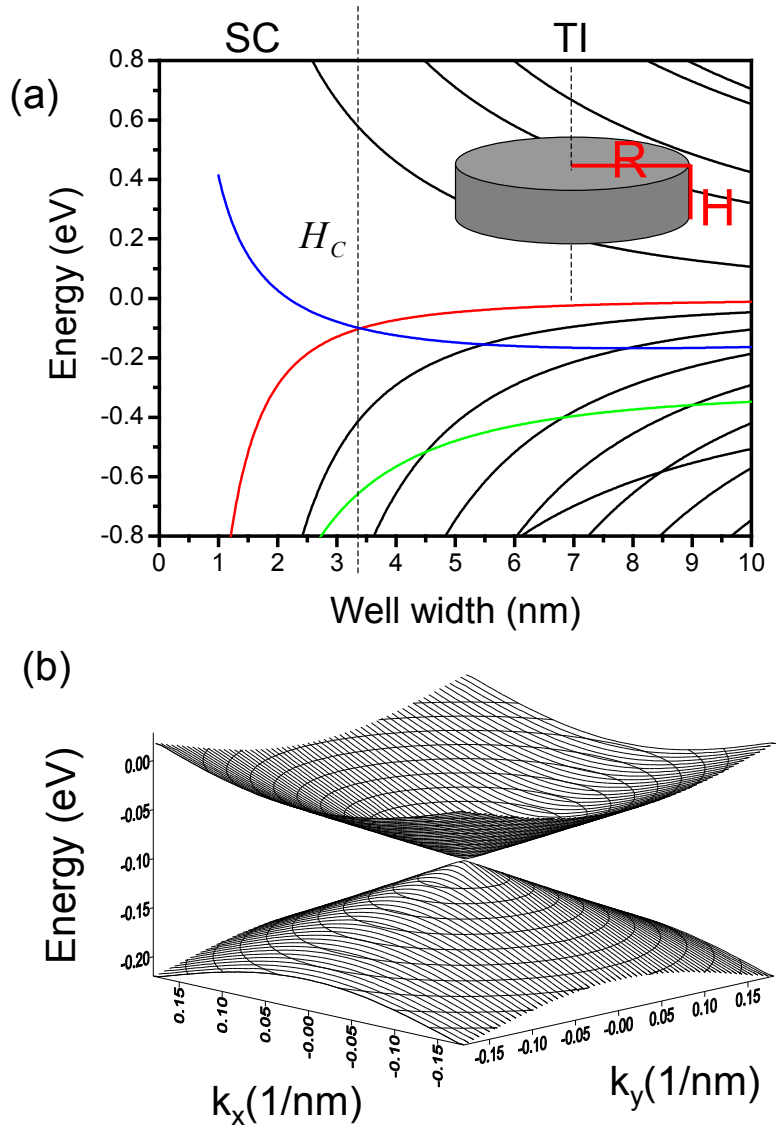}
\caption{(a) Subband edge energies in a HgTe quantum well at
  $k_x=k_y=0$ as a function of the well thickness $H$. 
  $H_C$ denotes the crossing between the
  lowest conduction (blue) and heavy-hole (red) subband edges, marking
  the transition between the normal semiconductor and the inverted
  bandgap material.
  Inset shows a schematic picture of the system composed of a single, 
  free-standing, disk-shaped  HgTe nanocrystal.
  (b) The in-plane dispersion of the conduction and heavy-hole
  subbands at the height $H_C$, showing the characteristic Dirac cone
  shape.}
\label{fig2n}
\end{figure}

We now turn to a microscopic description of a HgTe quantum dot based
on eight-band $\vec{k}\cdot\vec{p}$ theory developed for semiconductor
quantum
dots,\cite{michler-book,rego_hawrylak_prb1997,doty_climente_prl2009,climente_korkusinski_prb2008,korkusinski_hawrylak_prb2013}
HgTe quantum wells\cite{novik_pfeuffer_prb2005} and HgTe colloidal
nanocrystals.\cite{lhuillier_keuleyan_nanot2012,keuleyan_lhuillier_jacs2011,zhang_xia_jphysd2006}
We note that a simplified model of a strictly two-dimensional quantum
disk embedded in vacuum and described by a heavy hole and conduction
state has been recently studied by Chang and Lou.\cite{chang_lou_prl2011}
Specifically, we focus on a model colloidal HgTe quantum disk
with radius $R$ and height $H$ in vacuum, shown schematically in
the inset to Fig.~\ref{fig2n}(a).   
The single-particle energy levels and the corresponding wave functions
as a function of strain applied along the disk height are computed in
the eight-band $\vec{k}\cdot\vec{p}$ approach with the HgTe material
parameters, described in detail in the Methods section and taken from
Ref.~\onlinecite{novik_pfeuffer_prb2005}.

The eight-band $\vec{k}\cdot\vec{p}$ bulk Hamiltonian is written in
the basis of two conduction and six valence subband states. 
The eigenstates of the Hamiltonian are spinors with chirality up or
down (see Methods section).  
Each spinor carries conduction band, heavy hole, light hole and spin
split-off band components with a specific angular momenta for in-plane
motion and parity for vertical motion.
The chirality is a good quantum number and allows for the rigorous
classification of quantum states as observed experimentally by Doty
{\em et al.}\cite{doty_climente_prl2009,climente_korkusinski_prb2008}   
For example, we write the spinor with chirality ``up'' in the form
\begin{equation}
|\Uparrow,L\rangle = \left[
\begin{array}{c}
\sum\limits_{n=1}^{N} \sum\limits_{l=0}^{M} A_{nl}^{(1)}|n,m+1,2l+1\rangle |S\uparrow\rangle \\
\sum\limits_{n=1}^{N} \sum\limits_{l=0}^{M} A_{nl}^{(2)}|n,m+2,2l\rangle |S\downarrow\rangle \\
\sum\limits_{n=1}^{N} \sum\limits_{l=0}^{M} A_{nl}^{(3)}|n,m,2l+1\rangle \left|{3\over2}\right.,\left.+{3\over2}\right\rangle \\
\sum\limits_{n=1}^{N} \sum\limits_{l=0}^{M} A_{nl}^{(4)}|n,m+1,2l\rangle \left|{3\over2}\right.,\left.+{1\over2}\right\rangle \\
\sum\limits_{n=1}^{N} \sum\limits_{l=0}^{M} A_{nl}^{(5)}|n,m+2,2l+1\rangle \left|{3\over2}\right.,\left.-{1\over2}\right\rangle \\
\sum\limits_{n=1}^{N} \sum\limits_{l=0}^{M} A_{nl}^{(6)}|n,m+3,2l\rangle \left|{3\over2}\right.,\left.-{3\over2}\right\rangle \\
\sum\limits_{n=1}^{N} \sum\limits_{l=0}^{M} A_{nl}^{(7)}|n,m+1,2l\rangle \left|{1\over2}\right.,\left.+{1\over2}\right\rangle \\
\sum\limits_{n=1}^{N} \sum\limits_{l=0}^{M} A_{nl}^{(8)}|n,m+2,2l+1\rangle \left|{1\over2}\right.,\left.-{1\over2}\right\rangle \\
\end{array}
\right],
\label{spinor_up}
\end{equation}
where, $|n,m \rangle$ denote the in-plane basis functions with radial
node number $n$ and angular momentum $m$,  $|2l+1 \rangle$
($|2l\rangle$) describe the even (odd) trigonometric functions in the
$z$-direction, $\left|J\right.,\left.J_z\right\rangle$ are the subband
microscopic (Bloch) functions, and $A_{nl}^{(i)}$ are expansion
coefficients (see the Methods section).   
We see that the spinor contains all conduction band and valence band
states and that the different subband components enter with
different angular momentum and parity. 
This spinor is characterized by the total angular momentum quantum number 
$L=m+{3\over 2}$ which is a sum of the orbital angular momentum $m$
and the $z$-projection of the Bloch angular momentum $J$.
We have chosen to define the angular momenta of all spinor components
relative to the number $m$ of the heavy-hole component. 
In a similar way we define chirality ``down'' spinors.
The eight-band $\vec{k}\cdot\vec{p}$ Hamiltonian is expanded in the
basis of spinors and diagonalized to obtain coefficients $A$ as
explained in the Methods section. 
The wave functions vanish at the edge of the disk and any state
localized at the edge must be characterized by its effective radius $R^*$
and decay into the disk. 
Obtaining such localized states in terms of our basis functions is a
nontrivial numerical task.   


First we search for an optimal height $H$ of the HgTe quantum well
close to the transition from the normal semiconductor to the
topological insulator, i.e., from the normal to the inverted bandgap
regime. 
Figure~\ref{fig2n}(a) shows the energies of the quantum well
subbands in the vicinity of the critical thickness $H_C$, where
the conduction band edge belonging to the lowest subband (denoted in
blue) crosses the lowest-subband heavy-hole band edge (denoted in red). 
Thus, at sufficiently small well thicknesses $H$ we deal with the
normal semiconductor phase with the positive bandgap, while at larger
well thicknesses the material exhibits a bulk-like, inverted
bandstructure. 
The critical thickness for the infinite HgTe quantum well surrounded
by vacuum is found at $H_C=3.4$ nm.
Here, the in-plane dispersion corresponds to a single Dirac cone
as discussed and observed for HgTe/CdTe quantum
well.\cite{buettner_liu_natphys2011}
This Dirac cone is presented in Fig.~\ref{fig2n}(b).

\begin{figure}
\includegraphics[width=0.8\textwidth]{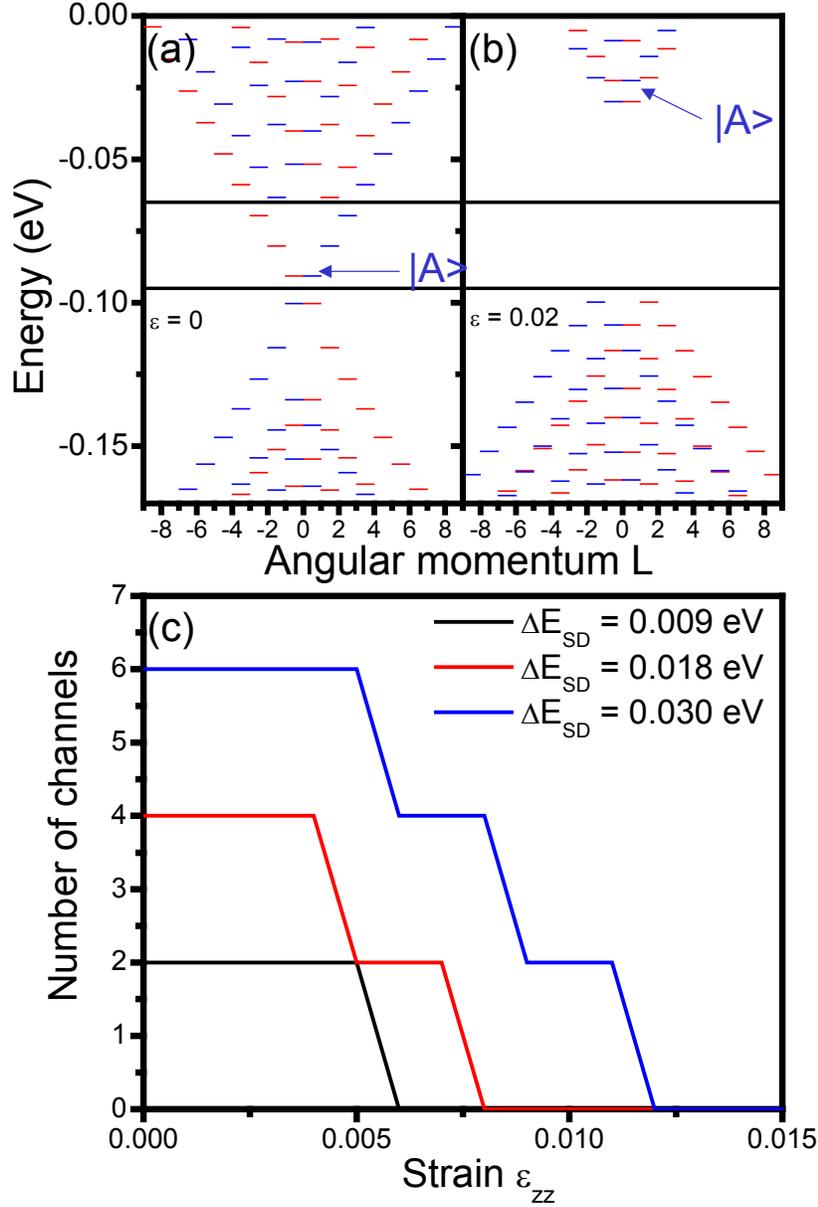}
\caption{ 
Single-particle states of the quantum disk of height $H=4$ nm
  and vertical strain $\varepsilon_{zz}=0$ (a), 
and $\varepsilon_{zz}=-0.02$ (b). 
The black horizontal lines define the conduction window established by
the Fermi levels of the leads of the electrical strain detector.
Red (blue) bars correspond to chirality ``up'' (``down'').
Panel (c) shows the current flowing through the sensor, measured in
terms of the number of conduction channels, as a function of the strain.
Different curves correspond to different widths of the conduction
window.} 
\label{fig3n}
\end{figure}

We now turn to a quantum disk with a height of $H=4$ nm which
corresponds to a quantum well in the TI regime. 
We unfold our spinors in linear combinations of $N$ Bessel functions
and $M$ sines for each subband, which results in the Hamiltonian
matrix of order of $K=8NM$ for each total angular momentum and
chirality channel.
In Fig.~\ref{fig3n}(a) we show the eigenenergies of the
quantum disk obtained without strain by diagonalizing the Hamiltonian
matrix with $N=40$ and $M=10$.
The levels, denoted by red and blue bars for chirality 
``up'' and ``down'', respectively, are plotted against the total
angular momentum $L$.
The states form degenerate Kramers pairs, one with chirality ``up'',
and the other with chirality ``down'', characterized by opposite total
angular momenta. 
We see the formation of two bands of edge states with quasi-linear
dispersion, the band with chirality ``up'' decreasing in energy as the
total angular momentum is increased, and the band with chirality
``down'' behaving oppositely.
Further, at energies higher than the edge bands we find the
``interior'' (non-edge) electron states, resembling those found in the
NS quantum dot.
The ``interior'' heavy-hole states form a ladder of levels below the
edge bands.

Fig.~\ref{fig3n}(b) illustrates the effect of strain included via the
Bir-Pikus Hamiltoninan on the single-particle energies. 
The compression $\varepsilon=-0.02$ along the disk height shifts
differently the electron and hole levels and tunes the HgTe bandgap,
such that its increase is analogous to moving from right to left in
Fig.~\ref{fig2n}(a). 
This results in a wider energy gap  in the quantum dot spectrum and
removal of the edge states.

We now discuss how the strain induced transition from a state with
edge states in the energy gap to the insulating state without such
states might be detected in transport. 
The two horizontal lines visible in Fig.~\ref{fig3n}(a) and (b)
denote the conduction window $\Delta E_{SD} = eV_{SD}$, where $V_{SD}$
is the source-drain voltage, defined by the Fermi levels of the left
(L) and right (R)leads.  
The number $M$ of edge states with the energy found within this window
contributes to the tunneling of electrons from one lead to the other,
and current $I=G_0 M T V_{SD}$ as discussed in the introduction.
With our choice of the $\Delta E_{SD}$, at zero strain we find six
states, while at $\varepsilon=-0.02$ the conduction window is empty
and the tunneling current cannot flow.
In Fig.~\ref{fig3n}(c) we show the effect of the width of the
source-drain voltage window on the number $M$
of available conduction channels as a function of the strain.
Different curves correspond to different source-drain voltages.
We see that a small conduction window contains fewer edge states,
exhibits a low critical strain and an abrupt transition from the
``on'' to ``off'' state.
This is the quantitative illustration of the central result of our
work presented in Fig.~\ref{fig1n}.
As the conduction window increases, the ``on'' current increases due
to the increased number of conduction channels, and the transition to
the ``off'' regime is less abrupt, consists of steps, and exhibits a
larger critical strain.
We expect, therefore, that the sensitivity and signal-to-noise ratio
can be significantly tuned simply by changing the source-drain voltage
of the leads.
The effect discussed here is to be contrasted with piezoelectric
sensors where induced charge is a linear function of applied
pressure,\cite{wang_zhou_nanolett2006,wang_advmat2012,wu_wen_science2013}
but has analogies to single-electron charge detection using
conductance steps in quantum point
contacts\cite{elzerman_hanson_nature2004,granger_taubert_natphys2012}
and pressure induced metal-insulator transitions in thin
films\cite{copel_kuroda_nl2013}. 
The magnitude of strain of few percent is comparable to strains
detected using piezoelectric semiconductor
nanowires.\cite{wang_advmat2012} 

\begin{figure}
\includegraphics[width=0.8\textwidth]{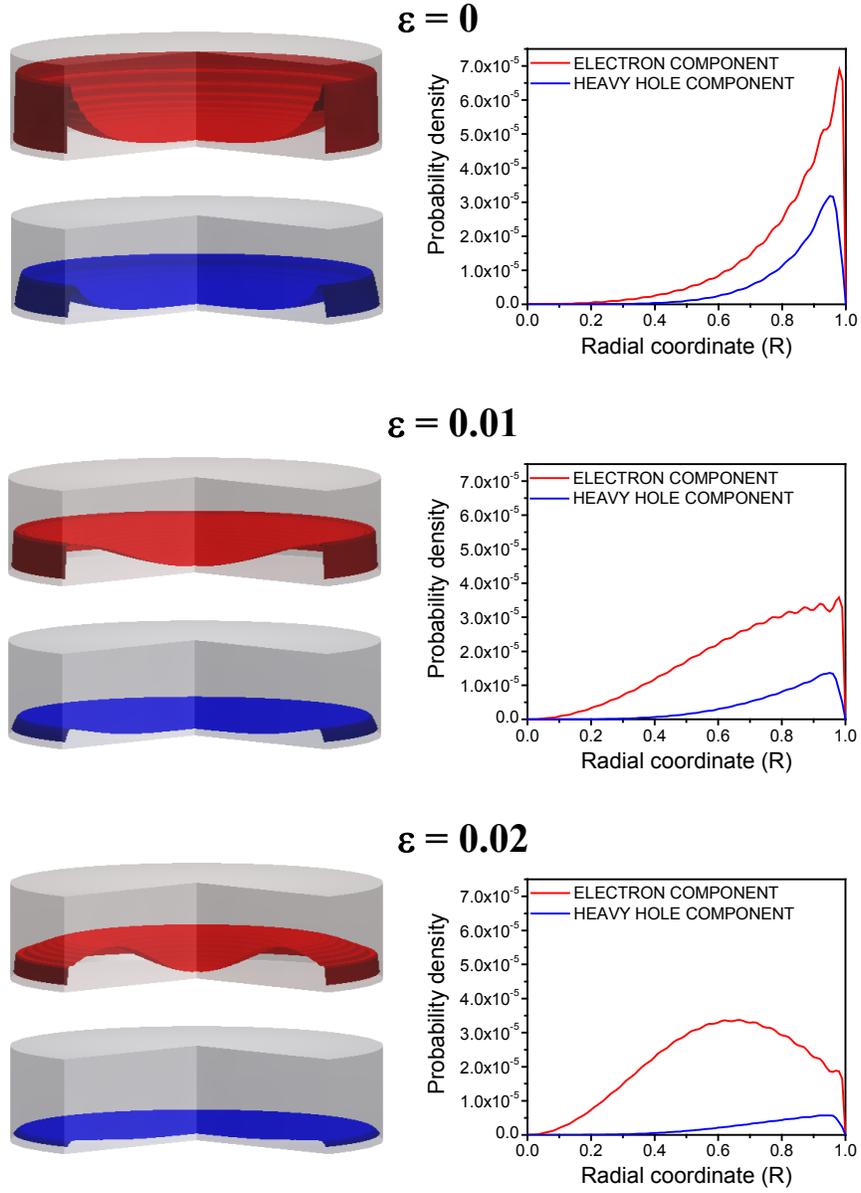}
\caption{ Probability densities of the state $|A\rangle$ from
  Fig.~\ref{fig3n} as a function of the radial 
  coordinate for different applied strains.
  The electron and heavy hole wave function components are shown in
  red and blue, respectively, in the form of three-dimensional plots
  in the left-hand column.  
  The radial plots along the $x$ axis are shown on the right with the
  same colour code.
  Densities for the strain $\varepsilon=0$, $0.01$, and $0.02$
  are shown in the top, middle, and bottom panels, respectively.}
\label{fig4n}
\end{figure}

We conclude our proposal by demonstrating that not only the number,
but also the nature of the conduction channels changes with the
strain.
In Fig.~\ref{fig4n} we show the radial dependence of the probability
densities of the state denoted in Fig.~\ref{fig3n} as $|A\rangle$.
The panels show, respectively, the electron (red) and heavy-hole
(blue) component of this state.
We see that as the strain is increased, the
probability density of the electronic component evolves from one
peaked at the edge to one peaked in the interior of the disk, while
the hole component drops to zero.
Thus, as expected, the states begin to resemble those of the NS
quantum disk.
As the probability density moves to the center, the tunneling from the
leads onto the conduction channels decreases exponentially.
Therefore, even though a state may be still within the conduction
window, its contact to leads (tunneling matrix element), and therefore
the tunneling current, decreases as a function of stress.

In conclusion, we present a microscopic theory of 
a HgTe quantum dot as a function of size and applied strain.
We demonstrated the existence of edge states and their removal
with applied strain.
We propose that this mechanism could be used to implement a nanoscale,
all-electrical, low-energy strain sensing device, in which the
presence of a strain beyond a threshold value could be detected
electrically as a collapse  of the tunneling current through the 
edge states of the HgTe quantum disk. 


\section*{Methods}

The essential physics of the TI edge states can only be captured if
one accounts for the strong mixing of the electron and hole
subbands.\cite{hasan_kane_rmp2010,qi_hughes_prb2008,konig_buhmann_jpsj2008,moore_nature2010,qi_zhang_rmp2011,zhou_lu_prl2008}
Here we analyze the single-particle properties of a HgTe disk, both
free-standing and embedded in a normal semiconductor material in the
eight-band $\vec{k}\cdot\vec{p}$ approach. 
This allows to relate its electronic properties directly to its
geometric and material properties.
Building on our previous work with NS quantum
wells\cite{rego_hawrylak_physe1998} and
dots,\cite{rego_hawrylak_prb1997,doty_climente_prl2009,climente_korkusinski_prb2008,korkusinski_hawrylak_prb2013}
we consider a HgTe quantum disk, shown schematically in
the inset to Fig.~\ref{fig2n}(a).
Here we discuss a free-standing nanocrystal, i.e., one whose surface
is modeled simply as an infinite potential barrier.
However, similar results were obtained for a HgTe quantum disk
embedded in the CdTe barrier material.
We take the radius of $R = 55$ nm and vary the height $H$ from $2$ nm
to about $10$ nm.
As already mentioned, in our calculations we
employ the eight-band $\vec{k}\cdot\vec{p}$ approach.
The Hamiltonian and relevant HgTe material parameters, described in
detail in the following, are taken from
Ref.~\onlinecite{novik_pfeuffer_prb2005}.
We note that a strictly two-dimensional model of the HgTe dot has been
considered in Ref.~\onlinecite{chang_lou_prl2011}.
In that approach, the effects of the disk height and subband mixing
were accounted for only through effective parameters of the electron
and the heavy hole in a two-band approach, which made its results
difficult to relate directly to realistic structure parameters.

The eight-band $\vec{k}\cdot\vec{p}$ bulk Hamiltonian is written in
the basis of two conduction and six valence subbands.
Denoting the spin of the quasiparticle with an arrow, 
$\sigma = \uparrow (\downarrow) = \pm 1/2$,
the Bloch basis set is chosen in the following form:
\begin{eqnarray}
\langle \vec{r} | S, +1/2\rangle & = & S\uparrow ,\\
\langle \vec{r} | S, -1/2\rangle & = & S\downarrow ,\\
\langle \vec{r} | 3/2, +3/2\rangle & = & (1/\sqrt{2})(X+iY)\uparrow ,\\
\langle \vec{r} | 3/2, +1/2\rangle & = & (1/\sqrt{6})[(X+iY)\downarrow
-2Z\uparrow] ,\\
\langle \vec{r} | 3/2, -1/2\rangle & = &-(1/\sqrt{6})[(X-iY)\uparrow
+2Z\downarrow] ,\\
\langle \vec{r} | 3/2, -3/2\rangle & = & -(1/\sqrt{2})(X-iY)\downarrow ,\\
\langle \vec{r} | 1/2, +1/2\rangle & = & (1/\sqrt{3})[(X+iY)\downarrow
+Z\uparrow] ,\\
\langle \vec{r} | 1/2, -1/2\rangle & = &(1/\sqrt{3})[(X-iY)\uparrow
-Z\downarrow] .
\end{eqnarray}
The total Bloch angular momentum for the electron subbands (the first
two states) is equal to the spin and is $1/2$.
The total Bloch angular momentum for the hole states is $3/2$ for the
heavy and light hole subbands, and $1/2$ for the spin-orbit split-off
subbands, and its projections are indicated by the second quantum
number in the ket.

The bulk Hamiltonian written in this basis takes the following form:
\begin{equation}
H = \left(
\begin{array}{cccccccc}
\hat{T}& 0 & -{1\over\sqrt{2}}P\hat{k}_{+} & 
\sqrt{2\over 3} P\hat{k}_z & {1\over\sqrt{6}} P\hat{k}_{-} & 
0 & - {1\over\sqrt{3}} P\hat{k}_z & - {1\over\sqrt{3}} P\hat{k}_{-} \\

0 & \hat{T} & 0 & - {1\over\sqrt{6}} P\hat{k}_{+} &
\sqrt{2\over 3} P\hat{k}_z & {1\over\sqrt{2}} P\hat{k}_{-} &
 -{1\over\sqrt{3}} P\hat{k}_{+} & {1\over\sqrt{3}} P\hat{k}_{z} \\

-{1\over \sqrt{2}} P\hat{k}_{-} & 0 & \hat{U}+\hat{V} & -\hat{S}_{-} &
\hat{R} & 0 & {1\over\sqrt{2}}\hat{S}_{-} & -\sqrt{2}\hat{R} \\

\sqrt{2\over 3}P\hat{k}_{z} & -{1\over\sqrt{6}}P\hat{k}_{-} & 
-\hat{S}_{-}^{+} & \hat{U}-\hat{V} & 0 & \hat{R} & \sqrt{2}\hat{V} & 
-\sqrt{3\over 2}\hat{S}_{-} \\

{1\over\sqrt{6}}P\hat{k}_{+} & \sqrt{2\over 3}P\hat{k}_{z} & \hat{R}^+
& 0 & \hat{U}-\hat{V} & \hat{S}_{-} & -\sqrt{3\over 2}\hat{S}_{+} & 
-\sqrt{2}\hat{V} \\

0 & {1\over\sqrt{2}}P\hat{k}_{+} & 0 & \hat{R}^+ & \hat{S}_+ & 
\hat{U}+\hat{V} & \sqrt{2}\hat{R}^+ & {1\over\sqrt{2}}\hat{S}_+ \\

-{1\over\sqrt{3}}P\hat{k}_z & -{1\over\sqrt{3}}P\hat{k}_- & 
{1\over\sqrt{2}}\hat{S}_-^+ & \sqrt{2}\hat{V} & 
-\sqrt{3\over 2}\hat{S}_+^+ & \sqrt{2}\hat{R} & \hat{U}-\Delta & 0 \\

-{1\over\sqrt{3}}P\hat{k}_+ & {1\over\sqrt{3}}P\hat{k}_z & 
-\sqrt{2}\hat{R}^+ & -\sqrt{3\over 2}\hat{S}_-^+ & -\sqrt{2}\hat{V} & 
{1\over\sqrt{2}}\hat{S}_+^+ & 0 & \hat{U}-\Delta \\
\end{array}
\right).
\label{eightbandhamil}
\end{equation}
The operators appearing in the above matrix are defined as:
\begin{eqnarray}
\hat{T} &=& E_c + \frac{\hbar^2}{2m_0} \left( \hat{k}_x^2
+ \hat{k}_y^2 + \hat{k}_z^2 \right), \\
\hat{U} &=& E_v - \frac{\hbar^2}{2m_0} \gamma_1 \left( \hat{k}_x^2
+ \hat{k}_y^2 + \hat{k}_z^2 \right), \\
\hat{V} &=& - \frac{\hbar^2}{2m_0} \gamma_2 \left( \hat{k}_x^2
+ \hat{k}_y^2 -2  \hat{k}_z^2 \right), \\
\hat{R} &=& \sqrt{3} \frac{\hbar^2}{2m_0} 
\left[ \gamma_2 \left(\hat{k}_x^2 - \hat{k}_y^2\right)
 -2i\gamma_3 \hat{k}_x\hat{k}_y \right], \\
\hat{S}_{\pm} & = & -2\sqrt{3} \frac{\hbar^2}{2m_0} \gamma_3
\hat{k}_{\pm} \hat{k}_z, 
\end{eqnarray}
and $\hat{k}_{\pm} = \hat{k}_x \pm i \hat{k}_y$.
Further, $m_0$ is the mass of a free electron and $\hbar$ is the Dirac
constant.
In what follows we take the following HgTe material
parameters:\cite{novik_pfeuffer_prb2005}
$\gamma_1 = 4.1$, $\gamma_2=0.5$, $\gamma_3 = 1.3$,
the spin-orbit splitting $\Delta = 1.08$ eV, and the bandgap
$E_g = E_c - E_v = -0.303$ eV, appropriate for our negative-badgap
material.
The conduction-valence subband coupling parameter $P$ can be
deduced from the Kane energy $E_P = 2m_0P^2/\hbar^2=18.8$ eV.
Lastly, we adopt the axial-symmetric approximation in which
$\bar{\gamma}=(\gamma_2+\gamma_3) / 2$ and the operator $\hat{R}$ can
be written in a simpler form 
$\hat{R} = \sqrt{3} \bar{\gamma} \frac{\hbar^2}{2m_0} \hat{k}_-^2$.
The Hamiltonian (\ref{eightbandhamil}) is applied to the
zero-dimensional quantum disk by the usual substitution
$\hat{k}_x = -i \partial/\partial x$, and analogously for
coordinates $y$ and $z$, and 
$\hat{k}_{\pm} = -i e^{\pm i\varphi}\left( \frac{\partial}{\partial r}
\pm i \frac{1}{r} \frac{\partial}{\partial \varphi} \right)$.
The natural choice of basis for the single-particle
states consists of eigenvectors of the single-band
Hamiltonian, 
\begin{equation}
\langle \vec{r} | n m l \rangle =
\frac{\sqrt{2}}{P} \frac{1}{|J_{m+1}(\alpha_m^n)|}
J_m\left( \alpha_m^n \frac{r}{P}\right) \times
\frac{1}{\sqrt{2\pi}} e^{im\varphi} \times
\sqrt{\frac{2}{W}}\sin\left( l\pi \frac{z}{W} \right),
\label{cylinderbasis}
\end{equation}
where $J_m$ is the Bessel function of $m$-th order, and
$\alpha_m^n$ is the n-th zero of that function.
The quantum numbers $n=0,1,\ldots$, 
$m= 0, \pm 1, \pm 2,\ldots$, and $l=1,2,\ldots$ are, respectively, the
nodal number and angular momentum of the in-plane function and
the vertical subband index.
Note that even (odd) values of $l$ correspond to odd (even) vertical
wave functions, with the origin of the coordinate system placed in the
center of the base of the disk. 
The functions by construction vanish on all the surfaces of the disk,
and therefore working with this basis we do not have to enforce any
additional boundary conditions.

The eigenvectors of the Hamiltonian (\ref{eightbandhamil}) are
sought in the form of eight-component spinors, each component being
a linear combination of single-subband functions.
However, the analysis of symmetries of the Hamiltonian allows to
discern two classes (``chiralities''), whose existence is related to
the Kramers degeneracy.
Owing to the fact that the operator $\hat{k}_+$ ($\hat{k}_-$)
increases (decreases) the envelope angular momentum by one unit, and
the operator $\hat{k}_z$ flips the symmetry of the vertical subband,
we write the spinor with chirality ``up'' in the form
presented in Eq.~\ref{spinor_up} (Refs.~\onlinecite{rego_hawrylak_prb1997,doty_climente_prl2009,climente_korkusinski_prb2008}).
This spinor is characterized by the total angular momentum quantum number
$L=m+{3\over 2}$ which is a sum of the orbital angular momentum $m$
and the $z$-projection of the Bloch angular momentum.
We have chosen to define the angular momenta of all spinor components
relative to the number $m$ of the heavy-hole component.
The spinor with chirality ``down'' takes the form
\begin{equation}
|\Downarrow,L\rangle = \left[
\begin{array}{c}
\sum\limits_{n=1}^{N} \sum\limits_{l=0}^{M} B_{nl}^{(1)}|n,m-2,2l\rangle |S\uparrow\rangle \\
\sum\limits_{n=1}^{N} \sum\limits_{l=0}^{M} B_{nl}^{(2)}|n,m-1,2l+1\rangle |S\downarrow\rangle \\
\sum\limits_{n=1}^{N} \sum\limits_{l=0}^{M} B_{nl}^{(3)}|n,m-3,2l\rangle \left|{3\over2}\right.,\left.+{3\over2}\right\rangle \\
\sum\limits_{n=1}^{N} \sum\limits_{l=0}^{M} B_{nl}^{(4)}|n,m-2,2l+1\rangle \left|{3\over2}\right.,\left.+{1\over2}\right\rangle \\
\sum\limits_{n=1}^{N} \sum\limits_{l=0}^{M} B_{nl}^{(5)}|n,m-1,2l\rangle \left|{3\over2}\right.,\left.-{1\over2}\right\rangle \\
\sum\limits_{n=1}^{N} \sum\limits_{l=0}^{M} B_{nl}^{(6)}|n,m,2l+1\rangle \left|{3\over2}\right.,\left.-{3\over2}\right\rangle \\
\sum\limits_{n=1}^{N} \sum\limits_{l=0}^{M} B_{nl}^{(7)}|n,m-2,2l+1\rangle \left|{1\over2}\right.,\left.+{1\over2}\right\rangle \\
\sum\limits_{n=1}^{N} \sum\limits_{l=0}^{M} B_{nl}^{(8)}|n,m-1,2l\rangle \left|{1\over2}\right.,\left.-{1\over2}\right\rangle \\
\end{array}
\right],
\label{spinor_down}
\end{equation}
where $N$ and $M$ define respectively the total number of radial and
vertical harmonics used in the calculation, and the total angular
momentum $L=m-{3\over 2}$. 
The problem is thus reduced to finding the sets of coefficients
$A_{nl}^{(i)}$ and $B_{nl}^{(i)}$ together with corresponding energy
eigenvalues by diagonalizing numerically the Hamiltonian matrix set up
in the above basis for each chirality, respectively.

Following Novik et al.\cite{novik_pfeuffer_prb2005} we account for the
strain effects by means of the Bir-Pikus Hamiltonian which is added to
the eight-band Hamiltonian, Eq.(~\ref{eightbandhamil}).
In the above basis, this Hamiltonian takes the following form:
\begin{equation}
\hat{H}_{BP} = \left[
\begin{array}{cccccccc}
\hat{t} & 0 & 0 & 0 & 0 & 0 & 0 & 0 \\
0 & \hat{t} & 0 & 0 & 0 & 0 & 0 & 0 \\
0 & 0 & \hat{u}+\hat{v} & \hat{s} & \hat{r} & 0 &
-\frac{1}{\sqrt2}\hat{s} & -\sqrt{2}\hat{r} \\
0 & 0 & \hat{s}^+ & \hat{u}-\hat{v} & 0 & \hat{r} & \sqrt{2}\hat{v} &
\sqrt{3\over 2}\hat{s} \\
0 & 0 & \hat{r} & 0 & \hat{u}-\hat{v} & -\hat{s} 
& \sqrt{3\over 2}\hat{s}^+ & -\sqrt{2}\hat{v} \\
0 & 0 & 0 & \hat{r}^+ & -\hat{s}^+ & \hat{u}+\hat{v} &
\sqrt{2}\hat{r}^+ & -\frac{1}{\sqrt2}\hat{s}^+\\
0 & 0 & -\frac{1}{\sqrt2} \hat{s}^+ & \sqrt{2}\hat{v} &
\sqrt{3\over 2}\hat{s} & \sqrt{2}\hat{r} & \hat{u} & 0 \\
0 & 0 & -\sqrt{2}\hat{r}^+ & \sqrt{3\over 2}\hat{s}^+ &
-\sqrt{2}\hat{v} & -\frac{1}{\sqrt2}\hat{s} & 0 & \hat{u}\\
\end{array}
\right],
\label{birpikus}
\end{equation}
with the following definitions of operators:
$\hat{t} = a_c
(\varepsilon_{xx}+\varepsilon_{yy}+\varepsilon_{zz})$,
$\hat{u} = a_v
(\varepsilon_{xx}+\varepsilon_{yy}+\varepsilon_{zz})$,
$\hat{v} = \frac{1}{2}b 
(\varepsilon_{xx}+\varepsilon_{yy}-2\varepsilon_{zz})$,
$\hat{s} = -d(\varepsilon_{xz}-i\varepsilon_{yz})$,
$\hat{r} = -\frac{\sqrt3}{2}b
(\varepsilon_{xx}-\varepsilon_{yy}) + id\varepsilon_{xy}$.
The strain defined by strain tensor matrix elements $\varepsilon_{ij}$
is translated into energy via deformation potentials $a_c$, $a_v$, $b$,
and $d$.
In this work we utilize the values given by van de
Walle\cite{vandewalle_prb1989}:
$a_c=-4.60$ eV, $a_v=-0.13$ eV, and $b=-1.15$ eV.
The value for the potential $d$ is not needed as we do not consider
any nondiagonal (shear) strain in the system.

In this work we consider a specific case of the disk being stressed in
the vertical direction (along its thickness).
We account for this case in the Bir-Pikus Hamiltonian,
Eq.~(\ref{birpikus}) by taking a nonzero strain tensor element
$\varepsilon_{zz}=\Delta H / H$, describing the relative change of the
dot height, and setting all other strain tensor matrix elements to
zero.
This approach neglects any buckling effects that might occur close to
the edges of the disk.
Under such simple strain, the only remaining nonzero elements of the
Bir-Pikus Hamiltonian are $\hat{t}$, which renormalizes the conduction
band edge, and $\hat{u}$ and $\hat{v}$, which influence the valence
subbands. 
Specifically, if the disk is stressed compressively
($\varepsilon_{zz}<0$), the conduction band edge is shifted towards
higher energies, since the relevant deformation potential $a_c<0$.
The heavy-hole band edge undergoes the shift of
$\hat{u}+\hat{v}=(a_v-b)\varepsilon_{xx}$.
Both valence-band deformation potentials are negative, but the
potential $b$ is an order of magnitude larger than $a_v$.
Under the negative strain, therefore, the heavy-hole subband edge is
shifted towards lower energies.
In total, the corrections to the conduction and heavy-hole band edges
contribute a positive term to the bandgap, i.e., tend to drive the
system from the inverted regime towards the normally gapped regime.
This property is the physical principle of operation of our strain sensor.

\section*{Author contributions}
M.K. and P.H. wrote the manuscript; M.K. prepared Figures 1-4.

\section*{Competing financial interests}
The Authors declare no competing financial interests in relation to
this work.

\newpage

\end{document}